\documentstyle[preprint,aps,epsf,floats]{revtex} 
\begin{document}

\draft
 
{\tighten
\preprint{\vbox{\hbox{CALT-68-2081}
                \hbox{CMU-HEP96-14}
                \hbox{UCSD-TH-96-26}
                \hbox{hep-ph/9610518} }}
 
\title{Corrections to the Bjorken and Voloshin sum rules}
 
\author{C.\ Glenn Boyd$\,^a$, Zoltan Ligeti$\,^b$,
  Ira Z.\ Rothstein$\,^c$, Mark B.\ Wise$\,^b$ }

\address{ \vbox{\vskip 0.truecm}
  $^a$Department of Physics, Carnegie Mellon University, 
    Pittsburgh, PA 15213 \\[6pt]
  $^b$California Institute of Technology, Pasadena, CA 91125 \\[6pt]
  $^c$Department of Physics, University of California at San Diego, 
    La Jolla, CA 92093 }

\maketitle 

\begin{abstract}%
We calculate near zero recoil the order $\alpha_s$ corrections to the Bjorken
and Voloshin sum rules that bound the $B\to D^{(*)}\,\ell\,\bar\nu$ form
factors.  These bounds are derived by relating the result of inserting a
complete set of physical states in a time ordered product of weak currents to
the operator product expansion.  The sum rules sum over physical states with
excitation energies less than a scale $\Delta$.  We find that the corrections 
to the Bjorken bound are moderate, while the Voloshin bound receives sizable
corrections enhanced by $\Delta/\Lambda_{\rm QCD}$.  With some assumptions, 
we find that the slope parameter for the form factor $h_{A_1}$ in 
$B\to D^*\,\ell\,\bar\nu$ decay satisfies $0.4\lesssim\rho_{A_1}^2\lesssim1.3$.

\end{abstract}

}%end tighten

\newpage

\section{Introduction}

It is possible to perform model independent extractions of some of the CKM
matrix elements and quark masses from exclusive and inclusive semileptonic $B$
meson decays via a systematic expansion in inverse powers of the heavy bottom
and charm quark masses.  The form factors in $B\to D^{(*)}\,\ell\,\bar\nu$
decays are related by heavy quark symmetry \cite{HQS} to the Isgur-Wise
function, $\xi(w=v\cdot v')$, where $v$ is the four-velocity of the $B$ and $v'$
is that of the $D^{(*)}$.  A model independent determination of $|V_{cb}|$ from
the differential decay rate ${\rm d}\Gamma(B\to D^*\,\ell\,\bar\nu)/{\rm d}w$
is made possible by the fact that $\xi$ is equal to unity at zero recoil ($w=1$)
\cite{HQS,NuWe,VoSi,Luke}.  Inclusive $B$ decay rates can be calculated by
performing an operator product expansion for the time ordered product of two
weak currents \cite{CGG}, allowing for a model independent determination of
$|V_{cb}|$ from the inclusive semileptonic decay rate, $\Gamma(B\to
X_c\,\ell\,\bar\nu)$.

The major theoretical uncertainties in the determination of $|V_{cb}|$ from
inclusive decays are due to the questionable convergence of the perturbative
corrections to the $b$ quark decay rate \cite{LSW}, and the uncertainties in the
$b$ and $c$ quark masses.  Uncertainties in the determination of $|V_{cb}|$ from
$B\to D^*\,\ell\,\bar\nu$ originate from order $\Lambda_{\rm QCD}^2/m_{c,b}^2$
corrections at zero recoil, and from extrapolating the form factors measured at
$w>1$ to $w=1$ (phase space vanishes at $w=1$).  The uncertainties in this
extrapolation would be reduced if the slope of the form factor at zero recoil
were known.

Sum rules have been derived that relate the exclusive decay form factors to the
inclusive decay rates.  The Bjorken sum rule \cite{Bj,IWsr} gives the bound
$\rho^2>1/4$, where $\rho^2$ is minus the slope of the Isgur-Wise function. 
Voloshin derived the upper bound
$\rho^2<1/4+(m_M-m_Q)/[2(m_{M_1}-m_M)]\simeq0.75$ \cite{Volo}, where
$m_M-m_Q=\bar\Lambda$ is the mass difference between the ground state heavy
meson and the heavy quark that it contains (up to corrections of order
$\Lambda_{\rm QCD}^2/m_{c,b}^2$), and $m_{M_1}-m_M$ is the mass of the first
excited meson state above the pseudoscalar-vector doublet.  

The most recent experimental data from CLEO \cite{CLEO} is $\rho_{B\to
D^*}^2=0.84\pm0.12\pm0.08$.  This might violate the above upper bound when
experimental uncertainties decrease.  The ALEPH \cite{ALEPH} result $\rho_{B\to
D^*}^2=0.29\pm0.18\pm0.12$ is significantly smaller, close to the above lower
bound.  The slope of the form factor $h_{A_1}$, which occurs in $B\to
D^*\,\ell\,\bar\nu$ decay, has also been studied by CLEO \cite{CLEOA1}. 
Central values for its slope parameter $\rho_{A_1}^2$ ranging between 0.91 and
1.53 have been obtained.  Thus, it is interesting to calculate the corrections
to the Bjorken and Voloshin bounds.

\section{Review of sum rules}

To derive the sum rules, we follow Refs.~\cite{BSUV,KLWG,GBIR}.  
Consider the time-ordered product
\begin{equation}\label{corrdef}
T_{\mu\nu} = {i\over2m_B}\, \int{\rm d}^4x\, e^{-iq\cdot x}\,
  \langle B \,|\, T\{J_\mu^\dagger(x),J_\nu(0)\} | B\, \rangle \,,
\end{equation}
where $J_\mu$ is a $b\to c$ axial or vector current, the $B$ states are at rest,
$\vec q$ is fixed, and $q_0=m_B-E_M-\epsilon$.  Here $E_M=\sqrt{m_M^2+|\vec
q\,|^2}$ is the minimal possible energy of the hadronic final
state\footnote{The ground state doublet of mesons have light degrees of freedom
with spin-parity $s_l^{\pi_l}=\frac12^-$.  We consider situations when only one
member of this doublet contributes.  It is this state that we denote by $M$.}
that can be created by the current $J_\mu$ with fixed $|\vec q\,|$.  With this
definition of $\epsilon$ in terms of the hadronic variables, the cut of 
$T_{\mu\nu}$ in the complex $\epsilon$ plane corresponding to physical states 
with a charm quark lies along $0<\epsilon<+\infty$.  It will be important that 
at the same value of $|\vec q\,|$ the cut at the parton level lies within the 
smaller region $\epsilon>\bar\Lambda(w-1)/w+{\cal O}(\Lambda_{\rm
QCD}^2/m_{c,b})$.  ($T_{\mu\nu}$ has another cut corresponding to physical 
states with two $b$ quarks and a $\bar c$ quark that lies between
$-2E_M>\epsilon>-\infty$.  This cut will not be important for our discussion.) 
To separate out specific hadronic form factors, one contracts the currents in
(\ref{corrdef}) with a suitably chosen four-vector $a$, yielding
\begin{equation}
a^{*\mu}\, T_{\mu\nu}(\epsilon)\, a^\nu = {1\over2m_B}\, 
  \sum_X\, (2\pi)^3\, \delta^3(\vec q+\vec p_X)\,
  {\langle B| J^\dagger\cdot a^* |X\rangle \langle X| J\cdot a |B\rangle \over
  E_X-E_M-\epsilon} + \ldots \,,
\end{equation}
where the ellipses denote the contribution from the cut corresponding to 
two $b$ quarks and a $\bar c$ quark.  The sum over $X$ includes the usual 
phase space factors, {\it i.e.}, ${\rm d}^3p/2E_X$ for each particle in 
the state $X$.

\begin{figure}[t]
\centerline{\epsfysize=9truecm \epsfbox{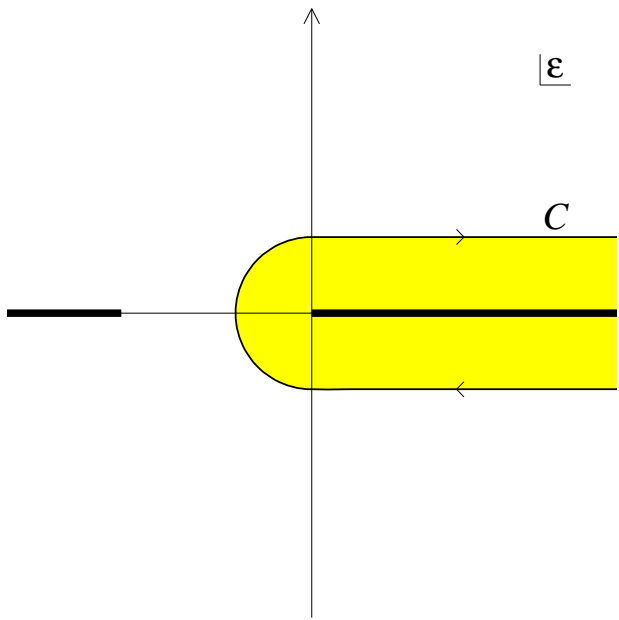}}
\caption[1]{The integration contour $C$ in the complex $\epsilon$ plane.
The cuts extend to ${\rm Re}\,\epsilon\to\pm\infty$. }
\end{figure}

While $T_{\mu\nu}(\epsilon)$ cannot be computed for arbitrary values of
$\epsilon$, its integrals with appropriate weight functions are calculable 
in perturbative QCD.  Consider integration of the product of a weight function
$W_\Delta(\epsilon)$ with $T_{\mu\nu}(\epsilon)$ along the contour $C$
surrounding the physical cut, shown in Fig.~1.  Assuming $W$ is analytic 
in the shaded region enclosed by this contour, we get 
\begin{equation}\label{zeroth}
{1\over2\pi i}\, \int_C {\rm d}\epsilon\, W_\Delta(\epsilon)\, 
  [a^{*\mu}\, T_{\mu\nu}(\epsilon)\, a^\nu]
= \sum_X\, W_\Delta(E_X-E_M)\, 
  (2\pi)^3\, \delta^3(\vec q + \vec p_X)\,
  {\Big|\langle X| J\cdot a |B\rangle\Big|^2\over2m_B}\, .
\end{equation}
The positivity of $|\langle X|J\cdot a|B\rangle|^2$ for all states $X$
gives an upper bound on the magnitude of form factors mediating $B$ decays
into the ground state doublet $M$.

The integral of the correlator weighted with $\epsilon\,W_\Delta(\epsilon)$
eliminates the contribution from the ground state doublet $X=M$, yielding
\begin{eqnarray}\label{first}
&& {1\over2\pi i}\, \int_C \epsilon\, {\rm d}\epsilon\, W_\Delta(\epsilon)\, 
  [a^{*\mu}\, T_{\mu\nu}(\epsilon)\, a^\nu] \nonumber\\
&& = \sum_{X\neq M}\, W_\Delta(E_X-E_M)\, 
  (2\pi)^3\, \delta^3(\vec q + \vec p_X)\, (E_X-E_M)\,
  {\Big|\langle X| J\cdot a |B\rangle\Big|^2\over2m_B}\,.
\end{eqnarray}
This can be turned into an upper bound on the contribution of excited states 
($X\neq M$) to the right-hand side of (\ref{zeroth}) by assuming that the
contribution of multi-hadron states is negligible below the first excited
meson state, $M_1$.  This is true in the large $N_c$ limit, and experimental
data available in the future on $B\to D^{(*)}\,\pi\,\ell\,\bar\nu$, 
{\it etc.}, decay rates can support (or oppose) the validity of this 
assumption.

Thus, there are upper and lower bounds
\begin{eqnarray}\label{bounds}
{1\over2\pi i}\, \int_C {\rm d}\epsilon\, W_\Delta(\epsilon)\, 
  [a^{*\mu}\, T_{\mu\nu}(\epsilon)\, a^\nu] 
&>& {\Big|\langle M| J\cdot a |B\rangle\Big|^2\over 4m_B\,E_M} \\
&>& {1\over2\pi i}\, \int_C {\rm d}\epsilon\, W_\Delta(\epsilon)\, 
  [a^{*\mu}\, T_{\mu\nu}(\epsilon)\, a^\nu]\,
  \bigg(1-{\epsilon\over E_{M_1}-E_M}\bigg) \,, \nonumber
\end{eqnarray}
where $E_{M_1}=\sqrt{m_{M_1}^2+|\vec q\,|^2}$.  
It should be emphasized that while this upper bound (which yields the Bjorken 
bound) is essentially model independent, the lower bound (which yields the 
Voloshin bound) relies on the above assumptions about the spectrum of the 
final state hadrons $X$.

Following \cite{KLWG}, we choose a set of weight functions 
\begin{equation}\label{weightfn}
W_\Delta^{(n)}(\epsilon) = {\Delta^{2n}\over\epsilon^{2n}+\Delta^{2n}} \,,
  \qquad (n=2,3,\ldots)
\end{equation}
that satisfy the following properties: ($i$) $W_\Delta$ is positive semidefinite
along the cut so that every term in the sum over $X$ on the hadron side of the
sum rule is non-negative; ($ii$) $W_\Delta(0)=1$; ($iii$) $W_\Delta$ is flat
near $\epsilon=0$; ($iv$) and $W_\Delta$ falls off rapidly to zero for
$\epsilon>\Delta$.  This choice of weight functions is motivated by the fact
that for values of $n$ of order unity all the poles of $W_\Delta^{(n)}$ lie at 
a distance of order $\Delta$ away from the physical cut.  As $n\to\infty$,
$W_\Delta^{(n)}$ approaches $\theta(\Delta-\epsilon)$ for $\epsilon>0$, which
corresponds to summing over all hadronic resonances up to excitation energy
$\Delta$ with equal weight.  In this limit the poles of $W_\Delta^{(n)}$
approach the cut, and the contour $C$ is forced to lie within a distance of
order $\Delta/n$ from the cut at $\epsilon=\Delta$.  In this case the
evaluation of the contour integrals using perturbative QCD relies on local
duality \cite{PQW} at the scale $\Delta$.  In the rest of this paper whenever
the weight function is not specified explicitly, we mean
$\theta(\Delta-\epsilon)$.  

The bounds in eq.~(\ref{bounds}) become weaker as $\Delta$ is increased. 
However, the scale $\Delta$ must be chosen large enough that the contour
integrals in eq.~(\ref{bounds}) can be performed using perturbative QCD,
allowing the evaluation of the Wilson coefficients of the operators that occur
in the operator product expansion for the time ordered product of currents.  In
practice this means that $\Delta$ must be greater than about $1$\,GeV.

\section{Bounds on the Isgur-Wise function}

The bounds stemming from eq.~(\ref{bounds}) are simplest to evaluate in the
heavy quark effective theory (HQET) \cite{eft}.  One may consider the vector
current in the effective theory, 
$V^\mu=\bar h^{(c)}_{v'}\,\gamma^\mu\,h^{(b)}_v$, and choose $a_\mu=v_\mu$, the
four-velocity of the $B$.  Instead of calculating the correlator itself, it is
simpler to compute its imaginary part given by the diagrams in Fig.~2.  In this
paper we focus on the region near zero recoil, and therefore we expand the
perturbative corrections to linear order in $(w-1)$.  

\begin{figure}[t]
\centerline{\epsfysize=1truein \epsfbox{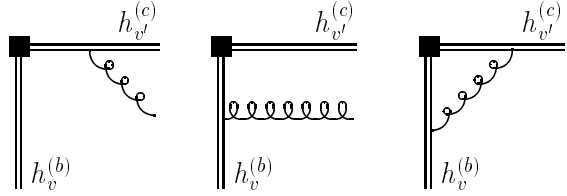}}
\caption[2]{Feynman diagrams that contribute to the order $\alpha_s$ 
corrections to the sum rules.  The heavy quark fields in the effective 
theory are denoted by $h^{(c,b)}$.  The black square indicates insertion 
of the $b\to c$ axial or vector current.}
\end{figure}

The ground state contribution is $\langle
M|V^\mu\,v_\mu|B\rangle=\sqrt{m_B\,m_M}\,(1+w)\,\xi(w)$, where $\xi(w)$ is the
Isgur-Wise function.  (Only the pseudoscalar member of the ground state doublet
contributes to this matrix element.)  In the $\overline{\rm MS}$ scheme, using
dimensional regularization and a finite gluon mass, $m_g$, the inclusive
expression for the correlator to order $\alpha_s$ reads 
\begin{eqnarray}\label{hqetcorr}
\frac1\pi\,{\rm Im}\,[T(\epsilon)] = {1+w\over2w}\, \bigg[ && \,
  \delta\bigg(\epsilon-\bar\Lambda\,{w-1\over w}\bigg)
  + (w-1)\, {8\alpha_s\over9\pi}\, \ln{m_g^2\over\mu^2}\, 
  \delta(\epsilon) \nonumber\\*
&& + (w-1)\, {8\alpha_s\over9\pi}\, {2\epsilon^2+m_g^2\over\epsilon^4}\,
  \sqrt{\epsilon^2-m_g^2}\, \theta(\epsilon-m_g) \bigg] + \ldots \,.
\end{eqnarray}
In eq.~(\ref{hqetcorr}) the terms proportional to delta functions come from
the charm quark final state, and the term proportional to
$\theta(\epsilon-m_g)$ arises from final states with a charm quark and a
single gluon.  The gluon mass is used to regulate an infrared divergence in the
integral of ${\rm Im}\,[T(\epsilon)]$ over $\epsilon$, which cancels (at order
$\alpha_s$) between these two types of final states.

Using eqs.~(\ref{zeroth}) and (\ref{first}) this implies  
the following sum rules
\begin{eqnarray}\label{hqetsr}
{1+w\over2w}\, \bigg[ 1 &+& (w-1)\, {8\alpha_s\over9\pi}\, 
  \bigg( \ln{4\Delta^2\over\mu^2} - \frac53\bigg) \bigg] =
  {(1+w)^2\over4w}\, |\xi(w)|^2 + \ldots \,, \nonumber\\*
{1+w\over2w}\, \bigg[ 1 &-& {\bar\Lambda\,(w-1)\over m_{M_1}-m_M}\, 
  \bigg( 1 + {16\alpha_s\over9\pi}\, {\Delta\over\bar\Lambda} \bigg) 
  \nonumber\\*
&+& (w-1)\, {8\alpha_s\over9\pi}\, \bigg( \ln{4\Delta^2\over\mu^2} 
  - \frac53 \bigg) \bigg] = {(1+w)^2\over4w}\, |\xi(w)|^2 - \ldots \,.
\end{eqnarray}
The ellipses in these equations denote positive terms whose first derivatives
at $w=1$ are also positive.  The reason for positivity of the first derivative
is that in the effective theory all excited state contributions must vanish at
$w=1$, and may therefore be written as $(w-1)$ times the square of some form
factor.  (Eq.~(\ref{hqetsr}) was previously obtained using a Wilson line 
approach to heavy quark interactions in Ref.~\cite{GrKo}.  
See also \cite{KoNe}.)

In eqs.~(\ref{hqetsr}), $\alpha_s$ is evaluated at the subtraction point $\mu$. 
This $\mu$-dependence on the left-hand side of eq.~(\ref{hqetsr}) is cancelled
by the $\mu$-dependence of the Isgur-Wise function, which we define in the
$\overline{\rm MS}$ scheme.  Differentiating with respect to $w$, we find the
following bounds on the slope parameter of the Isgur-Wise function,
$\rho^2=-{\rm d}\xi(w)/{\rm d}w|_{w=1}$,
\begin{equation}\label{hqetbounds}
\frac14 + {\bar\Lambda\over2(m_{M_1}-m_M)} + {4\alpha_s\over9\pi} 
  \bigg( {2\Delta\over m_{M_1}-m_M} + \frac53 - \ln{4\Delta^2\over\mu^2}\bigg)
> \rho^2(\mu) > \frac14 + {4\alpha_s\over9\pi} 
  \bigg( \frac53 - \ln{4\Delta^2\over\mu^2} \bigg) \,.
\end{equation}
Neglecting the order $\alpha_s$ corrections, these are precisely the Bjorken
and Voloshin bounds discussed in the introduction.  The upper bound on $\rho^2$
receives a perturbative correction of order $\alpha_s\,\Delta/\Lambda_{\rm
QCD}$, which is very large in the $\Delta\gg\Lambda_{\rm QCD}$
limit.\footnote{One is free to absorb all or part of this correction into a
redefinition of $\bar\Lambda$, $\bar\Lambda\to\bar\Lambda(\Delta)$, provided
one consistently reexpresses other formulae involving $\bar\Lambda$ in terms of
this new quantity.}  Note, however, that higher orders in perturbation theory
do not produce additional powers of $\Delta/\Lambda_{\rm QCD}$.  Similarly, sum
rules involving higher moments of ${\rm Im}[T(\epsilon)]$ \cite{BGSUV,BSUV}
will receive perturbative strong interaction corrections enhanced by more
powers of $\Delta/\Lambda_{\rm QCD}$.

The bounds on the slope of the Isgur-Wise function in eq.~(\ref{hqetbounds})
will have a perturbative series without large logarithms in its coefficients if
the subtraction point $\mu$ is chosen to be equal to $\Delta$.  Note that the
second term in the upper bound, proportional to $\bar\Lambda$, has a renormalon
ambiguity of order unity (since the heavy quark pole mass has a renormalon
ambiguity of order $\Lambda_{\rm QCD}$).  This is canceled by the ambiguity in
the perturbative series in $\alpha_s$ that multiplies $\Delta/(m_{M_1}-m_M)$,
the first term of which is presented in eq.~(\ref{hqetbounds}).

Using a weight function $W_\Delta^{(n)}(\epsilon)$ other than
$W_\Delta^{(\infty)}(\epsilon)=\theta(\Delta-\epsilon)$ does not affect the
lower (Bjorken) bound on $\rho^2$ given on the right-hand side of
eq.~(\ref{hqetbounds}).  Therefore, for the lower bound on $\rho^2$ (at order
$\alpha_s$), using the weight function $W_\Delta^{(\infty)}(\epsilon)$ does not 
rely on the assumption of local duality at the scale $\Delta$.  Such 
corrections, however, weaken the upper (Voloshin) bound by adding 
\begin{equation}
{4\alpha_s\over9\pi}\, {\Delta\over m_{M_1}-m_M}\, 
  \bigg( {\pi\over n\sin[\pi/(2n)]} - 2 \bigg) \,,
\end{equation}
to the left-hand side of eq.~(\ref{hqetbounds}). 
Numerical estimates of these corrections will be given later.

So far we have focused on the perturbative corrections to the coefficient of 
the lowest dimension operator, $\bar h^{(b)}_v\,h^{(b)}_v$, that occurs in the
operator product expansion for the time ordered product.  Higher dimension
operators are of the form $\bar h^{(b)}_v\,(v'\cdot D)^p\,h^{(b)}_v$.  These
yield corrections suppressed by powers of $\Lambda_{\rm QCD}/\Delta$ for weight
functions other than $W_\Delta^{(\infty)}(\epsilon)=\theta(\Delta-\epsilon)$.

A lower bound on $\rho^2$ including order $\alpha_s$ perturbative QCD
corrections was derived in Ref.~\cite{BGM}.  It corresponds to a weight
function given by the phase-space of $b$ decay, which is different from those
considered here.  The bound in \cite{BGM} appears stronger than that in
eq.~(\ref{hqetbounds}) because the weight function given by the phase-space
falls off faster with $\epsilon$.

To zeroth order in $\alpha_s$ and $\Lambda_{\rm QCD}/m_{c,b}$, the constraints
in the effective theory are identical with bounds on the slope of the measured
shape of the $B\to D^{(*)}\,\ell\,\bar\nu$ decay spectra.  However, at first
order in $\alpha_s$, one has to combine the above results with corrections that
originate from matching the full theory onto the heavy quark effective theory. 
[This will also eliminate the $\mu$-dependence from the bounds in
eq.~(\ref{hqetbounds}).]

\section{Differential decay rates}

We are interested in the form factors of semileptonic 
$B\to D^{(*)}\,\ell\,\bar\nu$ decays, defined as 
\begin{eqnarray}\label{ffdef}
{\langle D(v') | V^\mu | B(v)\rangle \over \sqrt{m_D\,m_B}} &=& 
  h_+(w)\,(v+v')^\mu + h_-(w)\,(v-v')^\mu \,, \nonumber\\*
{\langle D^*(v') | V^\mu | B(v)\rangle \over \sqrt{m_{D^*}\,m_B}} &=& 
  i\, h_V(w)\, \varepsilon^{\mu\nu\alpha\beta} 
  \epsilon^*_\nu v'_\alpha v_\beta\,, \\*
{\langle D^*(v') | A^\mu | B(v)\rangle \over \sqrt{m_{D^*}\,m_B}} &=& 
  h_{A_1}(w)\,(w+1)\,\epsilon^{*\mu} - h_{A_2}(w)\,(\epsilon^*\cdot v)\,v^\mu
  - h_{A_3}(w)\,(\epsilon^*\cdot v)\,v'^\mu \,. \nonumber
\end{eqnarray}
Here $V^\mu=\bar c\,\gamma^\mu\,b$ and $A^\mu=\bar c\,\gamma^\mu\gamma_5\,b$
are the vector and axial currents.  The kinematic variable $w$ is 
related to $q^2$ via $w=(m_B^2+m_{D^{(*)}}^2-q^2)/(2m_Bm_{D^{(*)}})$.
Up to corrections suppressed by powers of $\alpha_s(m_{c,b})$ and
$\Lambda_{\rm QCD}/m_{c,b}$, $h_-(w)=h_{A_2}(w)=0$ and
$h_+(w)=h_V(w)=h_{A_1}(w)=h_{A_3}(w)=\xi(w)$, where the Isgur-Wise function,
$\xi(w)$, is evaluated at a subtraction point around $m_{c,b}$.

Experimentally the differential decay rates are measured, and are usually quoted
in terms of the functions ${\cal F}_{B\to D^{(*)}}(w)$, defined below.  At tree
level, and without $\Lambda_{\rm QCD}/m_{c,b}$ corrections, these functions are
identical to the Isgur-Wise function, so their slopes are equal to that of
$\xi(w)$.  However, at order $\alpha_s$ additional corrections beyond those
calculated in eq.~(\ref{hqetbounds}) using the effective theory arise from
matching the full QCD onto the HQET.  Corrections suppressed by powers of
$\Lambda_{\rm QCD}/m_{c,b}$ arise from higher dimension operators in the HQET
Lagrangian, and from higher dimension current operators in the effective theory.
 
With the above definitions of the form factors, and $r^{(*)}=m_{D^{(*)}}/m_B$, 
the differential decay rates are
\begin{eqnarray}
{{\rm d}\Gamma(B\to D^*\,\ell\,\bar\nu)\over {\rm d}w} &=&
  {G_F^2\,m_B^5\over48\,\pi^3}\, r^{*3}\, (1-r^*)^2\, (w^2-1)^{1/2}\, (w+1)^2 
  \nonumber\\*
&\times& \bigg[1+{4w\over w+1}\,{1-2wr^*+r^{*2}\over(1-r^*)^2}\bigg]\, 
  |V_{cb}|^2\, |{\cal F}_{B\to D^*}(w)|^2 \,, \nonumber\\*
{{\rm d}\Gamma(B\to D\,\ell\,\bar\nu)\over {\rm d}w} &=&
  {G_F^2\,m_B^5\over48\,\pi^3}\, r^3\, (1+r)^2\, (w^2-1)^{3/2}\, |V_{cb}|^2\,
  |{\cal F}_{B\to D}(w)|^2 \,.
\end{eqnarray}
The functions ${\cal F}_{B\to D^*}$ and ${\cal F}_{B\to D}$ are given in terms 
of the form factors of the vector and axial currents defined in (\ref{ffdef}) as
\begin{eqnarray}\label{calf}
|{\cal F}_{B\to D^*}(w)|^2 &=& 
  \bigg[1+{4w\over w+1}\,{1-2wr^*+r^{*2}\over(1-r^*)^2}\bigg]^{-1}\, 
  \Bigg\{ {1-2wr^*+r^{*2}\over(1-r^*)^2}\, 2\,
  \bigg[ h_{A_1}^2(w) + {w-1\over w+1}\,h_V^2(w) \bigg] \nonumber\\*
&& \qquad\qquad\qquad + \bigg[ h_{A_1}(w) + {w-1\over 1-r^*}\, 
  \Big( h_{A_1}(w) - h_{A_3}(w) - r^*\,h_{A_2}(w) \Big) \bigg]^2 \Bigg\} \,, 
  \nonumber\\*
{\cal F}_{B\to D}(w) &=& h_+(w) - {1-r\over1+r}\, h_-(w) \,.
\end{eqnarray}
We define the ``physical" slope parameters, $\rho_{B\to D^*}^2$ and
$\rho_{B\to D}^2$, via
\begin{eqnarray}\label{physrho}
|{\cal F}_{B\to D^*}(w)| &=& |{\cal F}_{B\to D^*}(1)|\, 
  [1 - \rho_{B\to D^*}^2\, (w-1) + \ldots] \,, \nonumber\\*
|{\cal F}_{B\to D}(w)| &=& |{\cal F}_{B\to D}(1)|\, 
  [1 - \rho_{B\to D}^2\, (w-1) + \ldots ] \,.
\end{eqnarray}
Note that ${\cal F}_{B\to D^*}(1)=h_{A_1}(1)$.  Due to Luke's theorem 
\cite{Luke} $h_{A_1}(1)=\eta_A+{\cal O}(\Lambda_{\rm QCD}^2/m_{c,b}^2)$, while
${\cal F}_{B\to D}(1)=\eta_V+{\cal O}(\Lambda_{\rm QCD}/m_{c,b})$.  
The quantities $\eta_A$ and $\eta_V$ relate the axial and vector currents in 
the full theory of QCD to those in HQET at zero recoil.  

The order $\alpha_s$ corrections to the relationship between $\rho_{B\to
D^{(*)}}^2$ and the Isgur-Wise function can be computed model independently. 
We combine the results of the previous section with the order $\alpha_s$
matching corrections \cite{qcdcorr} taken from Ref.~\cite{physrep} to derive
bounds on the slope parameters.  Denoting $z=m_c/m_b$, and approximating
$r^{(*)}\simeq z$ in the order $\alpha_s$ corrections, the slope of the
Isgur-Wise function is related to that of ${\cal F}_{B\to D^{(*)}}$ via
\begin{equation}\label{rho2rel}
\rho_{B\to D^{(*)}}^2 = \rho^2(\mu) + {4\alpha_s\over9\pi} \ln{m_c^2\over\mu^2}
  + {\alpha_s\over\pi}\, \bigg( 
  \delta^{(\alpha_s)}_{B\to D^{(*)}} - {20\over27} \bigg)
  + {\bar\Lambda\over2m_c}\, \delta^{(1/m)}_{B\to D^{(*)}} \,. 
\end{equation}
Using eq.~(\ref{hqetbounds}) this implies the bounds
\begin{eqnarray}\label{hqetslope}
\rho_{B\to D^{(*)}}^2 &>& \frac14 
  + {4\alpha_s\over9\pi} \ln{m_c^2\over4\Delta^2}
  + {\alpha_s\over\pi}\, \delta^{(\alpha_s)}_{B\to D^{(*)}} 
  + {\bar\Lambda\over2m_c}\, \delta^{(1/m)}_{B\to D^{(*)}} \,, \\*
\rho_{B\to D^{(*)}}^2 &<& \frac14 
  + {\bar\Lambda\over2(m_{M_1}-m_M)}\, \bigg( 1 + 
  {16\alpha_s\over9\pi}\, {\Delta\over\bar\Lambda} \bigg) 
  + {4\alpha_s\over9\pi} \ln{m_c^2\over4\Delta^2}
  + {\alpha_s\over\pi}\, \delta^{(\alpha_s)}_{B\to D^{(*)}}
  + {\bar\Lambda\over2m_c}\, \delta^{(1/m)}_{B\to D^{(*)}} \,. \nonumber
\end{eqnarray}
The $\Delta$-independent part of the order $\alpha_s$ corrections is contained
in $\delta^{(\alpha_s)}_{B\to D^{(*)}}$, while $\delta^{(1/m)}_{B\to D^{(*)}}$ 
contains the order $\Lambda_{\rm QCD}/m_{c,b}$ corrections to 
$\rho_{B\to D^{(*)}}^2$.  We find
\begin{eqnarray}\label{deltaAs}
\delta^{(\alpha_s)}_{B\to D^*} &=&  
  {2\,(1-z)\,(11+2z+11z^2)+24\,(2-z+z^2)\,z\ln z 
  \over 27\,(1-z)^3} \,, \nonumber\\*
\delta^{(\alpha_s)}_{B\to D} &=& 
  {2\,(1-z)\,(23-34z+23z^2)+12\,(3-3z+2z^2)\,z\ln z \over 27\,(1-z)^3} \,. 
\end{eqnarray}

The corrections in $\delta^{(1/m)}_{B\to D^{(*)}}$ depend on the four
subleading Isgur-Wise functions \cite{Luke} that parametrize all first order
deviations from the infinite mass limit.  These can only be estimated at
present using model predictions.  Using the notation of \cite{physrep} we find
\begin{eqnarray}\label{delta1m}
\delta^{(1/m)}_{B\to D^*} &=& -2\chi_1'(1)+4\chi_3'(1)
  - z\, [2\chi_1'(1)-4\chi_2(1)+12\chi_3'(1)] \nonumber\\*
&& - \frac56\,(1+z) - \frac43\,\chi_2(1) - {1-2z+5z^2\over3(1-z)}\,\eta(1) \,, 
  \nonumber\\*
\delta^{(1/m)}_{B\to D} &=& - (1+z)\, [2\chi_1'(1)-4\chi_2(1)+12\chi_3'(1)]
  + {2(1-z)^2\over1+z}\, \eta'(1) \,.
\end{eqnarray}
Here prime denotes ${\rm d}/{\rm d}w$.

Note that the bounds in eqs.~(\ref{hqetslope}) do not rely on the assumption
that $m_{c,b}\gg\Delta$.  The bounds on the slope of the Isgur-Wise function in
HQET (where $m_{c,b}\rightarrow\infty$) hold as long as $\Delta$ is large
enough for perturbative QCD to be a valid way to calculate the contour
integrals in eq.~(\ref{bounds}).  The values for the charm and bottom quark
masses only arise in matching $\rho_{B\to D^{(*)}}^2$ onto the slope of the
Isgur-Wise function.

There are several scales that occur in the bounds on $\rho_{B\to D^{(*)}}^2$. 
In the limit $m_{c,b}\gg\Delta$ we know how to sum the large logarithms of the
ratio of scales $m_{c,b}/\Delta$ that occur in the perturbative corrections to
the bounds. First one performs the matching of $\rho_{B\to D^{(*)}}^2$ onto the
slope of the Isgur-Wise function at a scale around $m_{c,b}$ ({\it e.g.},
$\sqrt{m_cm_b}$), then one scales the Isgur-Wise function down from this
subtaction point to $\Delta$ using the anomalous dimension for the operator
$\bar h^{(c)}_{v'}\,\gamma^\mu\,h^{(b)}_v$.  Finally one applies the bound on
the slope of the Isgur-Wise function in eq.~(\ref{hqetbounds}) at the
subtraction point $\mu=\Delta$. However, since $\Delta$ must be greater than
$1\,$GeV, for the physical values of the charm and bottom quark masses
$m_{c,b}/\Delta$ is not very large. Consequently, we did not sum the leading
logarithms of this ratio using this renormalization group procedure in
eqs.~(\ref{hqetslope}).

\begin{table}[bt]
\begin{tabular}{c||cc|cc} 
&  \multicolumn{2}{c|}{Bjorken bound ($\rho_{B\to D^{(*)}}^2>\ldots$)}  
  &  \multicolumn{2}{c}{Voloshin bound ($\rho_{B\to D^{(*)}}^2<\ldots$)}  \\
&  $B\to D^*$  &  $B\to D$  &  $B\to D^*$  &  $B\to D$  \\  \hline   
~~$\Delta=1\,$GeV~~~~  &  0.24  &  0.33  &  0.95  &  1.05  \\
~~$\Delta=2\,$GeV~~~~  &  0.18  &  0.27  &  1.11  &  1.20
\end{tabular} \vskip6pt
\caption[]{Upper and lower bounds on $\rho_{B\to D^{(*)}}^2$, the slopes of the
functions ${\cal F}_{B\to D^{(*)}}(w)$ at zero recoil, that describe the shape
of the semileptonic $B\to D^{(*)}\,\ell\,\bar\nu$ decay spectrum.  The order
$\alpha_s$ corrections are included, while order $\Lambda_{\rm QCD}/m_{c,b}$
corrections are neglected.  To zeroth order in $\alpha_s$ the Bjorken bound 
is 0.25, while the Voloshin bound is 0.75.} 
\end{table}

To evaluate the bounds in eq.~(\ref{hqetslope}) we take\footnote{Unless
explicitly stated otherwise, we use these values throughout this paper.}
$\alpha_s=0.3$ (corresponding to a scale of about $2\,$GeV),
$\bar\Lambda=m_{M_1}-m_M=0.4\,$GeV, $m_c=1.4\,$GeV, and $m_b=4.8\,$GeV. 
However, since we neglect corrections of order $\alpha_s(\Lambda_{\rm
QCD}/m_{c,b})$, other values for the heavy quark masses, {\it e.g.}
$m_c\simeq\overline{m}_D=(m_D+3m_{D^*})/4=1.97\,$GeV and
$m_b\simeq\overline{m}_B=(m_B+3m_{B^*})/4=5.31\,$GeV, would be equally valid. 
In Table~I we show the Bjorken and Voloshin bounds at order $\alpha_s$ using
$\Delta=1\,$GeV and $\Delta=2\,$GeV, for the weight function
$W_\Delta^{(\infty)}=\theta(\Delta-\epsilon)$.  To obtain these numerical
results we neglected $\delta^{(1/m)}_{B\to D^{(*)}}$.  Using $W_\Delta^{(2)}$
instead of $W_\Delta^{(\infty)}$ does not affect the Bjorken bound, but weakens
the Voloshin bound by 0.02 for $\Delta=1\,$GeV, and by 0.05 for
$\Delta=2\,$GeV.  While the Bjorken bound only receives moderate corrections to
its tree-level value of 0.25, the corrections to the tree-level value of the
Voloshin bound, 0.75, are more sizable and strongly $\Delta$-dependent.  

To estimate the possible size of the order $\Lambda_{\rm QCD}/m_{c,b}$
corrections in eq.~(\ref{delta1m}), we use the QCD sum rule predictions for the
four subleading universal functions \cite{qcdsr}.  These give approximately
$\chi_1'(1)=0.3$, $\chi_2(1)=-0.04$, $\chi_3'(1)=0.02$, $\eta(1)=0.6$, and
$\eta'(1)=0$.  Note that the results for $\delta^{(1/m)}_{B\to D^{(*)}}$ depend
sensitively on $\chi_1'(1)$, which is only calculated to order $\alpha_s^0$ in
the framework of QCD sum rules; the other subleading form factors [$\chi_2(w)$,
$\chi_3(w)$, and $\eta(w)$] are computed to order $\alpha_s$.  We find that
these corrections reduce $\rho_{B\to D^*}^2$ by 0.3 and $\rho_{B\to D}^2$ by
0.2.  However, these values are model dependent, and the uncertainties are
large.

Using eq.~(\ref{rho2rel}) we find that the order $\alpha_s$ perturbative
corrections predict that $\rho_{B\to D}^2$ is about 0.09 larger than
$\rho_{B\to D^*}^2$.  This prediction is affected by order $\Lambda_{\rm
QCD}/m_{c,b}$ corrections [see eq.~(\ref{delta1m})], and therefore it is not
model independent.  However, the QCD sum rule results for the subleading
Isgur-Wise functions predict that $\rho_{B\to D}^2-\rho_{B\to D^*}^2$ is
further increased.  Therefore, an enhancement of $\rho_{B\to D}^2$ compared to
$\rho_{B\to D^*}^2$ by about $0.1-0.2$ seems quite likely, and a precise
measurement of this difference would test the predictions of Ref.~\cite{qcdsr}.
(Similar results were obtained in \cite{CaNe}.)

As the above model estimates for the order $\Lambda_{\rm QCD}/m_{c,b}$
corrections are fairly sizable, one should investigate whether more reliable
bounds can be derived in the full theory.  The motivation is that in the full
theory one can bound the magnitude of the physical form factors (equal to the
Isgur-Wise function plus order $\Lambda_{\rm QCD}/m_{c,b}$ corrections in
HQET).  In the next section we derive bounds on the $h_{A_1}(w)$ form
factor in the full theory.  For comparison, we give here the bounds on this
form factor in the effective theory approach.  The slope of $h_{A_1}(w)$ 
at $w=1$, $\rho_{A_1}^2$, satisfies a bound of the same form as that in
eq.~(\ref{hqetslope}).  The order $\alpha_s$ corrections to the bounds on
$\rho_{A_1}^2$ are
\begin{equation}\label{deltaAsA1}
\delta^{(\alpha_s)}_{A_1} = 
  {2\,(1-z)\,(17-4z+17z^2)+6\,(9-3z+4z^2)\,z\ln z \over 27\,(1-z)^3} \,,
\end{equation}
while the order $\Lambda_{\rm QCD}/m_{c,b}$ terms are
\begin{equation}\label{delta1mA1}
\delta^{(1/m)}_{A_1} = -2\chi_1'(1)+4\chi_3'(1)
  - z\, [2\chi_1'(1)-4\chi_2(1)+12\chi_3'(1)] - {1+z\over2} + z\,\eta(1) \,.
\end{equation}
Neglecting the corrections of order $\Lambda_{\rm QCD}/m_{c,b}$ this gives
$0.28<\rho_{A_1}^2<0.99$ for $\Delta=1\,$GeV and $0.22<\rho_{A_1}^2<1.15$ for 
$\Delta=2\,$GeV.

\section{Sum rules in the full theory}

In the full theory, bounds on the form factor $h_{A_1}(w)$ in eq.~(\ref{ffdef})
can be obtained from eq.~(\ref{bounds}) by taking the axial current and
choosing the four-vector $a$ such that $a\cdot v=a\cdot q=0$.  These bounds are
expected to be less model dependent at present than those derived in the
effective theory, as there is no uncertainty associated with the subleading
Isgur-Wise functions (they contribute at order $\Lambda_{\rm QCD}/m_{c,b}$ in
the effective theory approach).  On the other hand, the bounds derived in the
full theory receive calculable corrections proportional to powers of
$\Delta/m_{c,b}$ at order $\alpha_s$, which do not arise in the HQET approach.

The $\Delta$-dependent part of the corrections away from zero recoil are
straightforward to compute by considering the (finite) difference between the
bremsstrahlung graphs in the full theory and in HQET.  Since we want to keep
the full $w$ dependence at order $\Lambda_{\rm QCD}/m_{c,b}$, we need
to include
\begin{equation}
\bigg({1+w\over2w}\bigg)_{\rm parton} 
  = \bigg({1+w\over2w}\bigg)_{\rm hadron}\, 
  \bigg[1 - {(w-1)\over w^2}\, {\bar\Lambda\over m_c} + \ldots \bigg] \,.
\end{equation}
Neglecting terms of order $\alpha_s^2$, $\Lambda_{\rm QCD}^2/m_{c,b}^2$,
$\alpha_s(\Lambda_{\rm QCD}/m_{c,b})$, and $\alpha_s(w-1)^2$, we obtain the 
bounds
\begin{eqnarray}\label{fullbound}
{(1+w)^2\,|h_{A_1}(w)|^2\over4w} &<& {1+w\over2w}\, 
  \bigg[ 1-{(w-1)\over w^2}\, {\bar\Lambda\over m_c} \bigg]
  + {\alpha_s\over\pi} \Big[A+(w-1)X\Big] \,, \\*
{(1+w)^2\,|h_{A_1}(w)|^2\over4w} &>& {1+w\over2w}\, \bigg[ 1 - 
  {(w-1)\over w^2}\, {\bar\Lambda\over m_c} - {V\over E_{M_1}-E_M} \bigg]
  + {\alpha_s\over\pi} \Big[B+(w-1)Y\Big] \,. \nonumber
\end{eqnarray}
Here $A$, $B$, $X$, and $Y$ are functions of $m_c$, $m_b$, and $\Delta$.  
The term $V$ arises from the difference in the start of the parton and 
hadron cuts, and from matrix elements of dimension-5 terms in the operator 
product expansion.  It is simple to extract from \cite{BSUV,GBIR},
\begin{equation}
V = (w-1)\, \bigg( {\bar\Lambda\over w} 
  + {\bar\Lambda^2\over2m_c}\,{3-w\over w^3} 
  + {\lambda_1\over6m_c}\,{3+w\over w^3}
  + {\lambda_1+3\lambda_2\over3m_b}\,{1\over w} \bigg) + \ldots \,,
\end{equation}
where
\begin{eqnarray}
\lambda_1 &=& {1\over2m_B}\, \langle B(v) \,|\, \bar h_v^{(b)}\, (iD)^2\, 
  h_v^{(b)} \,|\, B(v)\rangle \,, \nonumber\\*
\lambda_2 &=& {1\over6m_B}\, \langle B(v) \,|\, \bar h_v^{(b)}\, {g\over2}\,
  \sigma_{\mu\nu}\, G^{\mu\nu}\, h_v^{(b)} \,|\, B(v)\rangle \,.
\end{eqnarray}
Since $E_{M_1}-E_M$ is of order $\Lambda_{\rm QCD}$, all terms in $V$ 
contribute at least of order $\Lambda_{\rm QCD}/m_{c,b}$ to the lower
bound on $|h_{A_1}(w)|^2$.  We cannot neglect these terms, as the main
motivation for considering the bounds in the full theory was to eliminate
the order $\Lambda_{\rm QCD}/m_{c,b}$ uncertainties related to the 
subleading Isgur-Wise functions in the HQET approach.

The function $A$ was computed in \cite{KLWG}\footnote{It was first computed
to order $\Delta^2/m_{c,b}^2$ in \cite{BSUV}.}, and $B$ was computed in 
\cite{GBIR}.  Denoting $d=\Delta/m_c$, the result is
\begin{eqnarray}\label{AB}
A &=& - 2\,\bigg({1+z\over1-z}\ln z +\frac83\bigg) 
  + {d\,(2+d)\, [2z^2(1+d)^2-(3+2z+z^2)] \over18\,(1+d)^2} 
  + {3+2z-z^2\over9} \ln(1+d) \,, \nonumber\\*
B &=& A - {\Delta\over m_{M_1}-m_M}\, \bigg[ 
  {(2+3d+2d^2)(9+6z-3z^2) +2d^2z^2(8+7d+2d^2) \over54\,(1+d)^2} \nonumber\\*
&& \phantom{A+{\Delta\over m_{M_1}-m_M}\,\bigg[} 
  - {3+2z-z^2\over9}\, {\ln(1+d)\over d} \bigg] \,.
\end{eqnarray}
We find that the coefficients $X$ and $Y$ are
\begin{eqnarray}
X &=& \bigg({1+z\over1-z}\ln z +\frac83\bigg) + \frac89 \ln(4d^2) 
  - 2\,\delta^{(\alpha_s)}_{A_1} - {d\,(16+42d+45d^2+16d^3) \over 9\,(1+d)^4}
  \nonumber\\*
&& - {2dz(20+52d+53d^2+18d^3) \over 45\,(1+d)^4} 
  + {dz^2(12+52d+71d^2+44d^3+10d^4) \over 45\,(1+d)^4} \nonumber\\*
&& - {80+12z^2\over45}\,\ln(1+d) \,, \nonumber\\
Y &=& X - {\bar\Lambda\,A\over m_{M_1}-m_M} 
  - {\Delta\over m_{M_1}-m_M} \bigg\{ {16\over9} 
  + {6-11d-62d^2-83d^3-32d^4 \over18\,(1+d)^4} \nonumber\\*
&& + {z(38+113d+122d^2+41d^3) \over45\,(1+d)^4} 
  - {z^2(34+119d+134d^2+49d^3-16d^4-10d^5) \over90(1+d)^4} \nonumber\\*
&& - {15+38z-17z^2\over45}\, {\ln(1+d)\over d} \bigg\} \,. 
\end{eqnarray}

\begin{figure}[tb]
\centerline{\epsfysize=9truecm \epsfbox{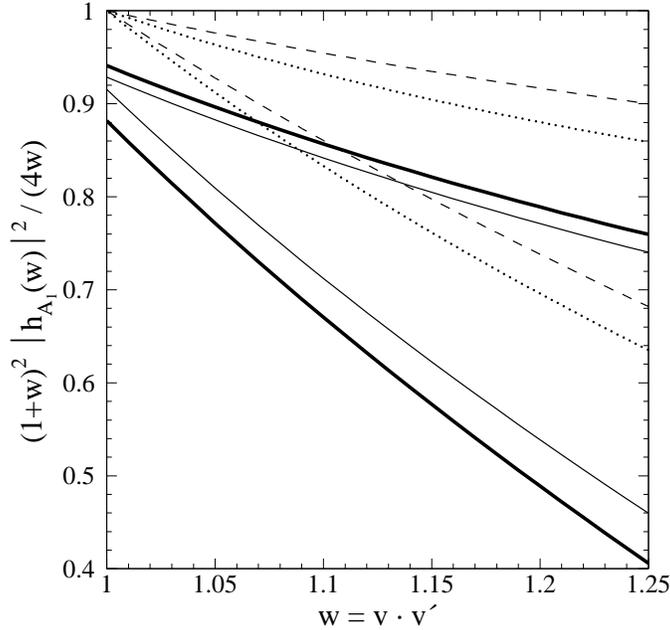}}
\caption[3]{Upper and lower bounds on $(1+w)^2\,|h_{A_1}(w)|^2/(4w)$.
The thin and thick solid curves are the bounds including effects of order
$\alpha_s$ and $\Lambda_{\rm QCD}/m_{c,b}$, and corespond to $\Delta=1\,$GeV 
and $2\,$GeV, respectively.  The dashed curves are the bounds neglecting order 
$\alpha_s$ and $\Lambda_{\rm QCD}/m_{c,b}$ corrections.  The dotted curves are 
the bounds neglecting order $\alpha_s$ but keeping $\Lambda_{\rm QCD}/m_{c,b}$ 
corrections.}
\end{figure}

In Fig.~3 we plot the upper and lower bounds on $(1+w)^2\,|h_{A_1}(w)|^2/(4w)$
over the region $1<w<1.25$, using eq.~(\ref{fullbound}).  Over this region of
$w$, corrections of order $\alpha_s(w-1)^2$ that we have not computed are
expected to be negligible.  The thin and thick solid curves correspond to the
choices $\Delta=1\,$GeV and $2\,$GeV, respectively.  The dashed curves show the
upper and lower bounds neglecting the order $\alpha_s$ and $\Lambda_{\rm
QCD}/m_{c,b}$ corrections.  The dotted curves show the upper and lower bounds
neglecting the order $\alpha_s$ corrections, but keeping the order
$\Lambda_{\rm QCD}/m_{c,b}$ terms.  The enhancement of the difference between
the upper and lower bounds, which is seen to increase with $w$, is dominated by
the perturbative corrections.  The reason for the somewhat larger than usual
deviation of 
\begin{equation}
|h_{A_1}(1)|^2 = \eta_A^2 = 1 - 
  2\,{\alpha_s\over\pi}\, \bigg({1+z\over1-z}\ln z +\frac83\bigg) + \ldots \,,
\end{equation} 
from unity in Fig.~3 is due to our choice of $\alpha_s=0.3$, which gives at 
order $\alpha_s$, $\eta_A=0.96$.

In plotting Fig.~3, we used $\lambda_1=-0.2\,{\rm GeV}^2$ and
$\lambda_2=0.12\,{\rm GeV}^2$.  While $\lambda_2$ is well-determined by the
$B^*-B$ mass splitting, the value of $\lambda_1$ is more uncertain.  Changing
$\lambda_1$ by $\pm0.2\,{\rm GeV}^2$ changes the lower bound at $w=1.25$ by
$\mp0.04$.  (At order $\Lambda_{\rm QCD}/m_{c,b}$ the value of $\lambda_1$ does
not affect the upper bound for all $w$, nor the lower bound at zero recoil.)  
We neglected the nonperturbative corrections of order $\Lambda_{\rm
QCD}^2/m_{c,b}^2$.  Such corrections to the lower bound involve matrix elements
of dimension-6 operators in the operator product expansion.  The order
$\Lambda_{\rm QCD}^2/m_{c,b}^2$ corrections to the upper bound [on the 
right-hand side of the first inequality in (\ref{fullbound})] are given by
\cite{BSUV,GBIR} 
\begin{equation}\label{1m2}
{w^2-1\over2w}\, {(\lambda_1-\lambda_2)w^2 - \bar\Lambda^2(3-2w^2)
  \over 2m_c^2\,w^4} + {\lambda_1-\lambda_2w^2\over4m_c^2\,w^5} +
  {\lambda_1+3\lambda_2\over12m_c^2\,w^3}\, (3w^2z^2+2z) \,.
\end{equation}
For the central values of $\bar\Lambda$ and $\lambda_1$ used throughout this
paper, these corrections affect the upper bounds by only about $-(0.02-0.03)$. 
The size of this correction is sensitive to the value of $\lambda_1$,
while it is largely independent of $\bar\Lambda$.  In Fig.~4 we plot the upper
(Bjorken) bound on $|h_{A_1}(w)|^2$ including the order $\Lambda_{\rm
QCD}^2/m_{c,b}^2$ corrections in eq.~(\ref{1m2}) for $\Delta=1\,$GeV and three
different values of $\lambda_1$, $\lambda_1=0$, $-0.2\,{\rm GeV}^2$, and
$-0.4\,{\rm GeV}^2$.  Changing $\bar\Lambda$ by $\pm0.1\,$GeV from the value we
used ({\it i.e.}, $0.4\,$GeV) affects the curves plotted in Fig.~4 by about 
$\mp0.01$.  If $\lambda_1$ is known more accurately in the future, this upper
bound may constrain $h_{A_1}(1)$ to be somewhat below $\eta_A$.

\begin{figure}[tb]
\centerline{\epsfysize=9truecm \epsfbox{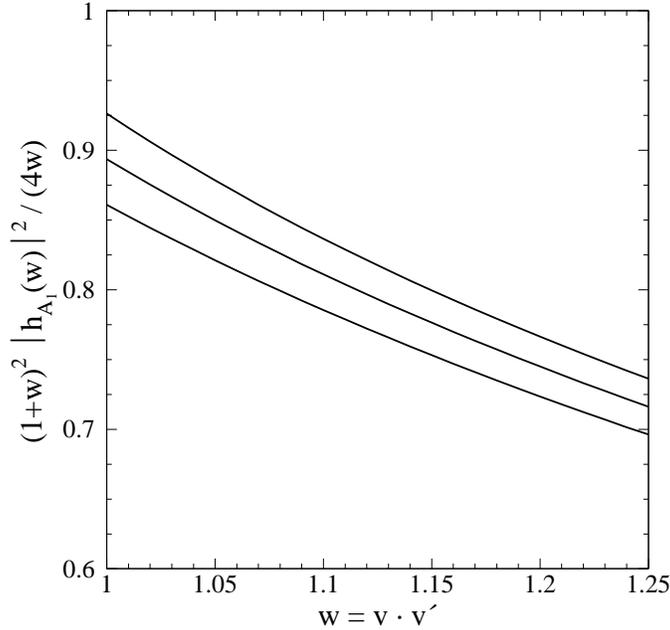}}
\caption[4]{Upper bound on $(1+w)^2\,|h_{A_1}(w)|^2/(4w)$ including order
$\Lambda_{\rm QCD}^2/m_{c,b}^2$ corrections for different values of 
$\lambda_1$.  The curves from above to below correspond to  
$\lambda_1=0$, $-0.2\,{\rm GeV}^2$, and $-0.4\,{\rm GeV}^2$, respectively.}
\end{figure}

A larger value of $\Delta$ increases our confidence in the validity of using
perturbative QCD to evaluate the time ordered product of weak currents.  In
going from $\Delta=2\,$GeV to $\Delta=3\,$GeV the upper bound (the upper thick
solid curve in Fig.~3) is increased by about $0.01$ fairly independently of $w$
over the region $1<w<1.25$, while the lower bound (the lower thick solid curve
in Fig.~3) is decreased by 0.06 at $w=1$ and by 0.07 at $w=1.25$.  Clearly, the
lower bound is considerably more sensitive to the choice of $\Delta$ than the
upper bound.  

Converting the bounds in eq.~(\ref{fullbound}) into constraints on the slope
parameter of the $h_{A_1}$ form factor at zero recoil, $\rho_{A_1}^2$, is not
straightforward.  The upper and lower bounds on $|h_{A_1}(w)|^2$ do not meet at
zero recoil, and therefore a bound on $\rho_{A_1}^2$ can only be derived with
some smoothness assumption.  We bound $\rho_{A_1}^2$ by assuming that
$h_{A_1}(w)$ is linear over the region $1<w<1.25$.  Then the plots in Fig.~3
(which neglect terms of order $\Lambda_{\rm QCD}^2/m_{c,b}^2$) imply the bounds
$0.44+4[h_{A_1}(1)-\eta_A]<\rho_{A_1}^2<1.19+4[h_{A_1}(1)-\eta_A]$ and
$0.39+4[h_{A_1}(1)-\eta_A]<\rho_{A_1}^2<1.36+4[h_{A_1}(1)-\eta_A]$ for
$\Delta=1\,$GeV and $2\,$GeV, respectively.  Recall that $h_{A_1}(1)-\eta_A$
is of order $\Lambda_{\rm QCD}^2/m_{c,b}^2$.  The increase in the lower bound
compared to the $1/4$ at zeroth order is mostly due to the terms proportional
to $\bar\Lambda/m_c$ in eqs.~(\ref{fullbound}), and hence it is sensitive to
the value of $\bar\Lambda$ we choose.  It is interesting that this correction
has the opposite sign than the QCD sum rule results in the effective theory,
which predicted that order $\Lambda_{\rm QCD}/m_{c,b}$ corrections lower the
values of the bounds.

The smoothness assumption used to derive bounds on the slope parameter
$\rho_{A_1}^2$ from Fig.~3 can be justified model independently using the
parameterization of Ref.\cite{BGL}.  This work also lets us extend the bounds
at small $w$ presented in Fig.~3 to larger values of $w$.  The $\Delta=2\,$GeV
bounds imply upper and lower bounds at $w=1.5$ of $0.86>h_{A_1}(1.5)>0.36$.  

The bounds presented in this section can also be used for unpolarized
$\Lambda_b\to\Lambda_c\,\ell\,\bar\nu$ decay if certain replacements are made.
The form factors for the matrix element of the axial current, $G_i$, are 
defined by
\begin{equation}
\langle \Lambda_c(v')| A^\mu | \Lambda_b(v)\rangle = \bar u(v')\, 
  [ G_1 \gamma^\mu + G_2 v^\mu + G_3 v'^\mu ]\,\gamma_5\, u(v) \,.
\end{equation}
Bounds on $G_1$ are obtained by replacing the left-hand side of 
eqs.~(\ref{fullbound}) by $G_1^2(w+1)/(2w)$.  The quantities $\bar\Lambda$,
$\lambda_1$, and $\lambda_2$ that appear on the right-hand side must now be
interpreted as arising from $\Lambda_b$ matrix elements.  They are simply 
related to the corresponding quantities in the $B$ meson case \cite{MaWi}, 
$\lambda_2(\Lambda_b)=0$, 
\begin{eqnarray}
\lambda_1(\Lambda_b) &=& \lambda_1(B) + 2m_c\, [(m_{\Lambda_b}-\overline{m}_B)
  - (m_{\Lambda_c}-\overline{m}_D)] / (1-z) \,,\nonumber\\*
\bar\Lambda(\Lambda_b) &=& \bar\Lambda(B) + m_{\Lambda_b}-\overline{m}_B
  + [\lambda_1(\Lambda_b)-\lambda_1(B)]/(2m_b) \,.
\end{eqnarray}

\section{Order $\alpha_{\lowercase{s}}^2\,\beta_0$ corrections at zero recoil}

At zero recoil we can rewrite the upper and lower bounds on the $h_{A_1}$
form factor in eq.~(\ref{fullbound}) as
\begin{eqnarray}\label{0recoil}
|h_{A_1}(1)|^2 &<& \eta_A^2 + {\alpha_s(\Delta)\over\pi}\, A_1 
  + {\alpha_s^2(\Delta)\over\pi^2}\, \beta_0\, A_2 \,, \\*
|h_{A_1}(1)|^2 &>& \eta_A^2 + {\alpha_s(\Delta)\over\pi}\, A_1 
  + {\alpha_s^2(\Delta)\over\pi^2}\, \beta_0\, A_2 -
  {\Delta\over m_{M_1}-m_M}\, \bigg[ {\alpha_s(\Delta)\over\pi}\, B_1 
  + {\alpha_s^2(\Delta)\over\pi^2}\, \beta_0\, B_2 \bigg] \,. \nonumber
\end{eqnarray}
These bounds also receive nonperturbative corrections of order 
$\Lambda_{\rm QCD}^2/m_{c,b}^2$.  Such corrections to the upper bound are
given by
\begin{equation}\label{uppernp}
-{\lambda_2\over m_c^2} + {\lambda_1+3\lambda_2\over4m_c^2}\,
  \bigg(1+\frac23\,z+z^2\bigg) \,.
\end{equation}
The order $\Lambda_{\rm QCD}^2/m_{c,b}^2$ corrections to the lower bound on
$|h_{A_1}(1)|^2$ depend on additional quantities (besides $\lambda_1$ and
$\lambda_2$) that parametrize matrix elements of dimension-6 operators
in the operator product expansion.

The corrections in $\eta_A$ in eq.~(\ref{0recoil}) arise, at the parton level,
from the final state $X=c$.  Except for an infrared renormalon ambiguity,
$\eta_A$ only depends on physics associated with the scales $m_{c,b}$.  It has
been calculated to order $\alpha_s^2\,\beta_0$ \cite{eta2}.  Explicitly,
\begin{equation}\label{etaA}
\eta_A = 1 - {\alpha_s(\sqrt{m_bm_c})\over\pi}\, \bigg({1+z\over1-z}\ln z
  + \frac83 \bigg) - {\alpha_s^2(\sqrt{m_bm_c})\over\pi^2}\, \beta_0\, 
  \frac5{24}\, \bigg({1+z\over1-z}\ln z + \frac{44}{15} \bigg) \,.
\end{equation}
The full order $\alpha_s^2$ expressions for $\eta_A$ and $\eta_V$ are also 
known approximately \cite{eta22}.  For $\eta_A$ the order $\alpha_s^2\,\beta_0$
terms dominate the $\alpha_s^2$ correction, while for $\eta_V$ they do not.

The corrections in $A_i$ and $B_i$ originate from final states $X$ that contain
a charm quark plus additional partons, {\it e.g.}, $c\,g$, $c\,\bar q\,q$, {\it
etc}.  These corrections are suppressed by powers of $\Delta/m_{c,b}$.  For
convenience, we evaluated the arguments of the strong couplings in
eqs.~(\ref{0recoil}) and (\ref{etaA}) at two different scales ($\sqrt{m_cm_b}$
for the series in $\eta_A$ and $\Delta$ for the other terms).  Of course, it is
possible to evaluate both series at the same scale using the QCD
$\beta$-function \cite{betaqcd}.  The functions $A_1$ and $B_1$ are given in
Sec.~V, $A_2$ was computed in Ref.~\cite{KLWG}, our result for $B_2$ is given
below.  These order $\alpha_s^2\,\beta_0$ corrections are relatively simple to
compute due to a relation between the $n_f$ dependent part of the order
$\alpha_s^2$ terms and the order $\alpha_s$ result with a finite gluon mass
\cite{SmVo}.  The calculation is simplest in the so-called $V$-scheme, but we
present the results in the usual $\overline{\rm MS}$ scheme.  To leading order
in $d=\Delta/m_c$
\begin{equation}
  \begin{array}{rclrcl}
A_1 &=& \displaystyle {d^2\over3}\, \bigg(1+\frac23\,z+z^2\bigg) \,, \qquad&
A_2 &=& \displaystyle \frac12\, \bigg({13\over6}-\ln2\bigg) A_1
  + {d^2\over15}\, \bigg(1+\frac43\,z+z^2\bigg) \,, \\*[6pt]
B_1 &=& \displaystyle \frac23\, A_1 \,, \qquad&
B_2 &=& \displaystyle \frac23\, A_2 - {d^2\over54}\, 
  \bigg(1+\frac23\,z+z^2\bigg) \,.
  \end{array}
\end{equation}
To all orders in $d$, $A_1$ and $B_1$ are given in eq.~(\ref{AB}), and $A_2$
and $B_2$ are computed numerically.  In Fig.~5 we plot the functions $A_1$,
$B_1$, $A_2$, and $B_2$ as a function of $\Delta$.  The dotted curve is $A_1$,
the dash-dotted curve is $B_1$, the dashed curve is $A_2$, and the solid curve
is $B_2$.

\begin{figure}[t]
\centerline{\epsfysize=9truecm \epsfbox{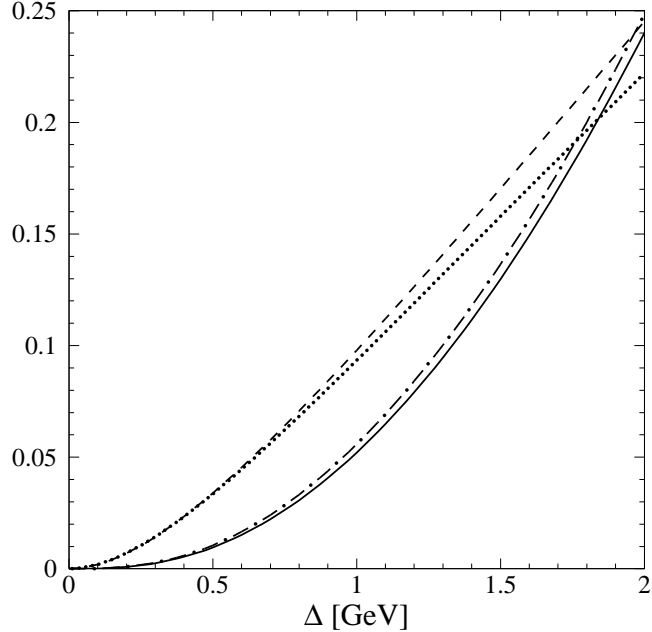} }
\caption[5]{The functions $A_1$, $B_1$, $A_2$, and $B_2$ defined in 
eq.~(\ref{0recoil}) as a function of $\Delta$.  
The dotted curve is $A_1$, the dash-dotted curve is $B_1$, the dashed curve 
is $A_2$, and the solid curve is $B_2$. }
\end{figure}

On the right-hand sides of eqs.~(\ref{0recoil}) the renormalon ambiguity of
order $\Lambda_{\rm QCD}^2/m_{c,b}^2$ in $\eta_A^2$ is cancelled by that in the
series $A_i$.  Therefore, in considering the size of the $\alpha_s^2$
corrections relative to those of order $\alpha_s$, it seems reasonable for the
upper bound to take all terms on the right-hand side of eqs.~(\ref{0recoil})
together \cite{MS}.  Although $A_2$ is approximately as big as $A_1$ (over the
region $\Delta<2\,$GeV), since the order $\alpha_s^2\,\beta_0$ term in
$\eta_A^2$ partially cancels against $A_2$, the perturbative expansion for the
upper bound seems reasonably well-behaved.\footnote{Ref.~\cite{KLWG} was less
certain that the upper bound on $|h_{A_1}(1)|^2$ has a well-behaved
perturbative expansion.  This is mostly due to the fact that in \cite{KLWG} the
behavior of the perturbative series arising from the terms $A_i$ (which contain
all the $\Delta$-dependence) was discussed without combining it with that from
$\eta_A^2$.}  For example, with $\Lambda_{\rm QCD}=200\,$MeV and
$\Delta=1\,$GeV, one finds neglecting terms of order $\Lambda_{\rm
QCD}^2/m_{c,b}^2$ that
\begin{eqnarray}
|h_{A_1}(1)|^2 < 1 &-& 0.073 - 0.019 \nonumber\\*
  &+& 0.013 + 0.017 \nonumber\\*
= 1 &-& 0.060 - 0.002 \,.
\end{eqnarray}
The last two terms in the first line are the order $\alpha_s$ and
$\alpha_s^2\,\beta_0$ corrections to $\eta_A^2$, while the second line contains
the terms proportional to $A_1$ and $A_2$.  For values of $\Delta$ between
$1\,$GeV and $5\,$GeV the cancellation of order $\alpha_s^2\,\beta_0$ terms
persists.

For the lower bound, the order $\alpha_s^2\beta_0$ term in the perturbative
expansion of the term proportional to $\Delta/(m_{M_1}-m_M)$, which originates
from the first moment of the time ordered product of weak currents, is about as
large as the order $\alpha_s$ term over the region $\Delta<5\,$GeV.  With 
$\Lambda_{\rm QCD}=200\,$MeV and $\Delta=1\,$GeV, the lower bound in 
eq.~(\ref{fullbound}) is
\begin{eqnarray}
|h_{A_1}(1)|^2 > 1 &-& 0.060 - 0.002 \nonumber\\*
  &-& 0.019 - 0.022 \nonumber\\*
= 1 &-& 0.079 - 0.024 \,.
\end{eqnarray}
The terms in the first line arise from $\eta_A^2$ and the series $A_i$, while
the second line contains the terms proportional to $B_1$ and $B_2$.  Note that
the coefficient of the $B_i$ terms depend on the mass of the first excited 
state.

\section{Conclusions}

In this paper we studied corrections to the Bjorken and Voloshin sum rules on
form factors of semileptonic $B\to D^{(*)}\,\ell\,\bar\nu$ decays.  In the
heavy quark effective theory we derived upper and lower bounds on the
Isgur-Wise function, and on its slope at zero recoil.  Matching the full theory
onto HQET, we translated the bounds in the effective theory into bounds on
${\cal F}_{B\to D^{(*)}}$, the shape of the measured $B\to D^{(*)}$ spectrum. 
The results in Table~I show that while the corrections to the Bjorken bound
(lower bound on $\rho_{B\to D^{(*)}}^2$) are small, the corrections to the
Voloshin bound (upper bound on $\rho_{B\to D^{(*)}}^2$) are sizable.  The
reason is that perturbative corrections to the Voloshin bound are enhanced by
$\Delta/\bar\Lambda$.  Therefore, even if experimental data would settle around
$\rho_{B\to D^{(*)}}$ slightly above unity, that would still not be a problem
for the theory to accommodate.

The bounds derived in Sec.~IV are affected at order $\Lambda_{\rm QCD}/m_{c,b}$
by corrections that are parametrized by four subleading universal functions,
and are not known at present model independently.  Therefore, we also studied
the sum rule constraints on the $h_{A_1}(w)$ form factor in the full theory. 
In this approach, all $\Lambda_{\rm QCD}/m_{cb}$ corrections to the sum rules
are parametrized by the three matrix elements, $\bar\Lambda$, $\lambda_1$, and
$\lambda_2$.  Bounds on the $h_{A_1}$ form factor are shown in Fig.~3, and with
some assumptions we found that its slope parameter satisfies
$0.4\lesssim\rho_{A_1}^2\lesssim1.3$.

These bounds on $h_{A_1}$ in the full theory of QCD can also be related to
bounds on $\rho_{B\to D^{(*)}}^2$.  Using eqs.~(\ref{deltaAs}) and
(\ref{deltaAsA1}) we find that the order $\alpha_s$ corrections imply
$\rho_{B\to D^*}^2=\rho_{A_1}^2-0.04$ and $\rho_{B\to D}^2=\rho_{A_1}^2+0.05$. 
These relations receive order $\Lambda_{\rm QCD}/m_{c,b}$ corrections.  Due to
heavy quark spin symmetry the order $\Lambda_{\rm QCD}/m_{c,b}$ difference
between $\rho_{B\to D^{(*)}}^2$ and $\rho_{A_1}^2$ is independent of the
subleading Isgur-Wise function $\chi_1$.  Then eqs.~(\ref{delta1m}) and
(\ref{delta1mA1}), together with the QCD sum rule predictions for the
subleading Isgur-Wise functions, imply that order $\Lambda_{\rm QCD}/m_{c,b}$
terms reduce $\rho_{B\to D^*}^2-\rho_{A_1}^2$ by 0.11 and reduce $\rho_{B\to
D}^2-\rho_{A_1}^2$ by 0.01.  (The uncertainty in these predictions will be
reduced if the form factor ratio $R_2$ can be measured precisely.)

One of the largest uncertainties in the sum rule predictions (especially in the
Voloshin bound) is related to the numerical values of $\bar\Lambda$ and
$m_{M_1}-m_M$.  In this paper we used $m_{M_1}-m_M=\bar\Lambda=0.4\,$GeV,
motivated by the experimentally measured $D_1-D^*$ mass difference, and by the
extraction of $\bar\Lambda$ in \cite{GKLW}.  However, the uncertainties in this
determination of $\bar\Lambda$ are sizable, and precise experimental data on
other inclusive processes is needed to extract the value of this quantity more
reliably \cite{pheno}.  Taking $m_{M_1}-m_M=m_{D_1}-m_{D^*}$ may also be
misleading.  Besides the possibility of sizable decay rates into non-resonant
final states \cite{NI} (discussed in Sec.~II), there is probably a doublet
lighter than the $\{D_1,D_2^*\}$ heavy quark spin symmetry doublet, that is of
order $100\,$MeV or more broad.  The spin-parity of the light degrees of
freedom in the $\{D_1,D_2^*\}$ doublet is $s_l^{\pi_l}=\frac32^+$, and so the
$D_1$ is $1^+$, while the $D_2^*$ is $2^+$.  However, light degrees of freedom
with $s_l^{\pi_l}=\frac12^+$ yield a doublet of $0^+$ and $1^+$ states.  These
can decay into $D^{(*)}\pi$ in an $s$-wave, and so they should be much broader
than the $\{D_1,D_2^*\}$ that can only decay in a $d$-wave.  (An $s$-wave decay
amplitude for the $D_1$ is allowed by angular momentum conservation, but it is
forbidden by heavy quark spin symmetry \cite{IWprl}.)  A reduction in
$m_{M_1}-m_M$ would further weaken the Voloshin bound.

Some improvements in this paper are possible.  We focused on the region near
zero recoil because it is important for the extraction of $|V_{cb}|$.  It would
be straightforward to calculate the terms in eq.~(\ref{fullbound}) of order
$\alpha_s(w-1)^2$.  Then Fig.~3 could be extended over the full kinematic range
$1<w<1.5$.  Some uncertainty in the sum rules arises from the order
$\alpha_s^2$ corrections.  The part of these corrections proportional to the
one-loop $\beta$-function have been computed at zero recoil.  For the part of
the lower bound involving the first moment of the time ordered product of weak
currents these corrections are as big as the order $\alpha_s$ corrections,
unless $\Delta$ is quite large.  It should be possible to compute the
$\alpha_s^2\,\beta_0$ corrections away from zero recoil, and also to the bounds
on $\rho^2_{B\rightarrow D^{(*)}}$ considered in Section III.

\acknowledgements
We would like to thank B. Grinstein for critical reading of the manuscript.
C.G.B. thanks L. Wolfenstein for stimulating conversations, and the
nuclear theory group at University of Washington for their hospitality during 
the completion of this work.  
This work was supported in part by the U.S.\ Dept.\ of Energy under Grant no.\
DE-FG03-92-ER~40701, DE-FG02-91ER40682, and DOE-FG03-90ER40546.  

{\tighten
  
}%end tighten

\end{document}